# Search for anti-quark nuggets via their interaction with the LHC beam

K. Zioutas[a], A. Zhitnitsky[b], C. Zamantzas[c], Y. K. Semertzidis[d], O. M. Ruimi[e,f,o], K. Ozbozduman[g], M. Maroudas[h,*], A. Kryemadhi[i], M. Karuza[j], D. Horns[h], A. Gougas[k], S. Cetin[l], G. Cantatore[m], and D. Budker[f,n,o]

[a] *University of Patras, Physics Department, Patras, Greece*

[b] *University of British Columbia, Department of Physics and Astronomy, Vancouver, Canada*

[c] *European Organization for Nuclear Research (CERN), Geneva, Switzerland*

[d] *IBS / KAIST, Daejeon, Korea*

[e] *Hebrew University of Jerusalem, Racah Institute of Physics, Jerusalem, Israel*

[f] *Johannes Gutenberg University Mainz, Helmholtz Institute, Mainz, Germany*

[g] *Boğaziçi University, Physics Department, Istanbul, Türkiye*

[h] *University of Hamburg, Hamburg, Germany*

[i] *Messiah University, Mechanicsburg, PA, USA*

[j] *University of Rijeka, Rijeka, Croatia*

[k] *Pylon Group, Athens, Greece*

[l] *Istinye University, Istanbul, Türkiye*

[m] *University and INFN Trieste, Trieste, Italy*

[n] *University of California, Department of Physics, Berkeley, California, USA*

[o] *Helmholtz-Institut, GSI Helmholtzzentrum fur Schwerionenforschung, Mainz, Germany*

**E-mail: marios.maroudas@cern.ch*

## Abstract

Anti-quark nuggets (AQNs) have been suggested to solve the dark matter (DM) and the missing antimatter problem in the universe and have been proposed as an explanation of various observations. Their size is in the μm range and their density is about equal to the nuclear density with an expected flux of about 0.4 / km$^2$ / year. For the typical velocity of DM constituents (~250 km/s), the solar system bodies act as highly performing gravitational lenses. Here we assume that DM streams or clusters are impinging, e.g., on the Earth, as it was worked out for DM axions and Weakly Interacting Massive Particles (WIMPs). Interestingly, in the LHC beam, unforeseen beam losses are triggered by so-called Unidentified Falling Objects (UFOs), which are believed to be constituted of dust particles with a size in the μm range and a density of several orders of magnitude lower than AQNs. Prezeau suggested that streaming DM constituents incident on the Earth should result in jet-like structures ("hairs") exiting the Earth, or a kind of caustics. Such ideas open novel directions in the search for DM. This work suggests a new analysis of the UFO results at the Large Hadron Collider (LHC), assuming that they are eventually, at least partly, due to AQNs. Firstly, a reanalysis of the existing data from the 4000 beam monitors since the beginning of the LHC is proposed, arguing that dust and AQNs should behave differently. The feasibility of this idea has been discussed with CERN accelerator people and potential collaborators.

## Introduction

The search for the direct detection of particles from the dark sector failed so far despite the worldwide efforts to unravel its nature. DM and dark energy remain two of the biggest questions in physics. All we know so far is that DM interacts gravitationally with the normal matter of the visible Universe and that our Universe is expanding in an accelerating manner. The applied techniques utilize a variety of processes like the recoiling nuclei by the celebrated WIMPs, the conversion of axions to real photons inside magnetic fields *à la* Sikivie, but also more novel techniques including highly sensitive accelerometers networks including accelerometers based on atomic interferometry [1], and the axion echo method [2, 3]. Most of these measurements are in operation and certainly address a variety of candidate constituents from the dark sector we live in. Given the multifaceted forms of visible matter, it is reasonable to assume that the dominating dark sector also consists of an eventually larger variety of constituents giving rise to an even larger complexity. With this proposal, we address the AQNs as they have been theoretically proposed by A. Zhitnitsky in 2003 [4]. In fact, AQNs were suggested to solve two major problems in physics: the DM problem and the apparent charge asymmetry of our universe. Namely, do exist planets, stars, etc made of anti-matter? Of note, matter and anti-matter should in principle both be created in equal amounts in the early Universe, and the question is, where is all the anti-matter? Also, this challenging question addresses the tentative concept that the AQNs are based on.

With the present proposal, we focus on the direct detection of DM in the form of AQNs using existing LHC data. Of course, the same search is suggestive for other accelerators. After all, if DM consists of many components, their variety might fit into specific accelerator types. Finally, we stress the parasitic form of the suggested direct search for DM using machine data stored from the LHC operation since its start.

## The proposal

AQNs became the possible cause of various observations occurring in our vicinity or outer space. Their size is in the μm range with nuclear density. Interestingly, in the LHC beam has been discussed the appearance of dust particles with size also in the μm range (dubbed UFOs, Unidentified Falling Objects). Their density is several orders of magnitude lower than that of the introduced AQNs suggesting to reconsider the investigations of the UFOs, having in mind also the AQNs and their potential interaction with a proton beam like that of the LHC machine at CERN. Of course, also other accelerators or colliders are of potential interest and would be considered. Of note, a relevant parameter is the binding energy (~60 MeV) of the ~$10^{25}$ anti-quarks which make up the main constituents of the AQNs.

The expected flux of AQNs is about 0.4 / $km^2$ / year. Streaming DM offers a viable common scenario following gravitational focusing by the solar system bodies [5,6]. This fits in as the underlying process behind the solar cycle, which was the first signature following a planetary dependency. R. Wolf suspected already in 1859, a planetary association for the 11-year solar cycle coinciding (within ~7%) with the 11.8-year orbital periodicity of Jupiter. The challenge, ever since is to find a remote planetary impact, beyond the extremely feeble planetary gravity (tidal force). As the typical velocity of DM constituents is ~250 km/s, this makes all solar system bodies highly performing gravitational lenses for DM streams or clusters [7]. Remarkably, also the inner Earth mass distribution is expected to perform as a gravitational lens giving rise to flux enhancements by a factor up to $10^8$, following Prezeau [8] and Sofue [9]. In addition, DM streams incident on the Earth could exit from the Earth or other solar system bodies as jet-like structures ("hairs") or caustics-like [8]. While the direct search for DM has failed to observe constituents from the dark sector, such ideas suggest novel concepts for detecting constituents from the dark sector we live in.

The UFO studies at the LHC encourage us to reconsider the achieved conclusions assuming AQNs instead. In short, for the suggested search of the theoretically proposed AQNs in LHC, existing data from ~4000 beam monitors, since the beginning of the LHC, need to be re-analyzed. The monitors are distributed along the two beam pipes with the two counter-rotating beams inside. Interestingly, the data taken by the ~4000 monitors are stored and could be reanalyzed. Of note, dust and AQNs should interact with the multi-TeV proton bunches differently. Also, their time dependence should be different and can be used for identification purposes. For example, UFOs do not have an external origin, e.g. exo-solar system. By contrast, DM particles can have a kind of "memory" associated with their cosmic origin. This implies a possible sidereal time dependence that leads to characteristic rhythms, which could be used for the possible signal identification. Therefore, it seems worth the effort to undertake such a reanalysis of the existing LHC data.

It is reasonable to assume that more accelerators could apply this reasoning aiming to unravel a new signature from the dark sector, which remained so far well hidden. LHC has the potential to become pioneering in a new type of direct detection of DM, and this parasitically to its rich physics schedule. After all, LHC and the future FCC have as flagship also the detection of DM.

## Conclusion

We propose here that the 4000 detectors of the LHC beam monitoring system can be used to search for AQNs, which are theoretically proposed constituents of DM. The interaction of both LHC proton beams may interact with one single nugget (eventually several times since each LHC bunch consists of about $10^{11}$ protons). The binding energy of the nugget's constituents is ~60 MeV and therefore not a barrier for the LHC beam energy.

In future, depending on the gained experience from the reanalysis of the stored data from the 4000 beam monitors, a more appropriate monitoring system can be designed and set, up or the data-saving software of the existing monitors can be adjusted accordingly.

Finally, while we have addressed the LHC for the present proposal, the same reasoning could also be applied to any other accelerator types. Of note, it is reasoning to assume that DM consists of several different constituents, each fitting in better a specific accelerator.

## Acknowledgments

We wish to thank Louis Wahlkiers and Marco Buzio for the very interesting discussions we had during the preparation of this proposal at the very beginning. MM acknowledges funding by the Deutsche Forschungsgemeinschaft (DFG, German Research Foundation) under Germany's Excellence Strategy– EXC 2121 "Quantum Universe"– 390833306, and through the DFG funds for major instrumentation grant DFG INST 152/8241."